\documentclass{aa}

\def\msun{{\rm M}_\odot}
\def\mbar{M_{\rm B}}
\def\msol{{\rm M}_\odot}
\def\msolyr{\rm M_\odot/yr}
\def\rms{r_{\rm ms}}
\def\ltot{l_{\rm tot}}

\def\pd{\dot P}
\def\pms{P_{\rm ms}}
\def\ltot{l_{\rm tot}}
\def\lmag{l_{\rm m}}
\def\rms{r_{\rm ms}}
\def\rmag{r_{\rm m}}
\def\rcor{r_{\rm c}}
\def\medd{\dot M_{\rm Edd}}
\def\md{\dot M}

\usepackage{graphicx}
\usepackage{natbib}
\usepackage{amssymb,amsmath}

\begin{document}

\title{Progenitor neutron stars of the lightest and heaviest millisecond pulsars}
\author{ M.~Fortin\inst{1,2,3} \and M. Bejger\inst{1} \and P. Haensel\inst{1}\and J. L. Zdunik\inst{1}}
\institute{N. Copernicus Astronomical Center, Polish
           Academy of Sciences, Bartycka 18, PL-00-716 Warszawa, Poland
\and
LUTh, UMR 8102 du CNRS, Observatoire de Paris, F-92195 Meudon Cedex, France
\and
Istituto Nazionale di Fisica Nucleare - Sezione di Roma, P.le Aldo Moro 2, 00185 Roma, Italy\\
{\tt fortin@camk.edu.pl, bejger@camk.edu.pl, haensel@camk.edu.pl, jlz@camk.edu.pl}}
\offprints{M. Fortin}
\date{Received xxx Accepted xxx}

\abstract{The recent mass measurements of two binary millisecond pulsars,
PSR J1614$-$2230 and PSR J0751+1807 with a mass $M=1.97\pm
0.04\;\msun$ and $M= 1.26 \pm 0.14 \;\msun$, respectively,
indicate a wide range of masses for such objects and possibly also
a broad spectrum of masses of neutron stars born in
core-collapse supernov\ae.}
{Starting from the zero-age main sequence binary stage, we aim at
inferring the birth masses of PSR J1614$-$2230 and PSR J0751+1807
by taking the differences in the evolutionary stages preceding
their formation into account.}
{Using simulations for the evolution of binary stars, we reconstruct
the evolutionary tracks leading to the formation of PSR J1614$-$2230 and PSR J0751+1807. We analyze
in detail the spin evolution due to the accretion of matter from a disk in the intermediate-mass/low-mass
X-ray binary. We consider two equations of state of dense matter, one for purely nucleonic matter and the other one including a
high-density softening due to the appearance of hyperons. Stationary and axisymmetric stellar configurations in general relativity are used, together with a recent magnetic torque model and
observationally-motivated laws for the decay of magnetic field.}
{The estimated birth mass of the neutron stars PSR J0751+1807 and PSR J1614$-$2230
could be as low as $1.0\;\msun$ and as high as $1.9\;\msun$, respectively. These values depend
weakly on the equation of state and the assumed model for the magnetic field and its accretion-induced decay.}
{The masses of progenitor neutron stars of recycled pulsars
span a broad interval from $1.0\;\msun$ to $1.9\;\msun$. Including the effect of a slow
Roche-lobe detachment phase, which
could be relevant for PSR J0751+1807, would make the lower mass limit
even lower. A realistic theory for core-collapse supernov\ae\ should
account for this wide range of mass.}


\titlerunning{Progenitors of PSR J1614$-$2230 and PSR J0751+1807}
\authorrunning{Fortin et al.}
\maketitle

%
\section{Introduction}
\label{sect:introd}
Millisecond radio pulsars (defined here as those with a
spin period $P<10\;$ms) have several unique properties that make them
very interesting objects to study, both observationally and theoretically. They are
the most rapid stellar rotators with a spin frequency $f=1/P$ up to 716 Hz \citep{Hessels2006}. Their spin periods
are extremely stable with a typical period increase, owing to the spin angular momentum
loss associated with magneto-dipole radiation: $\dot{P}\sim 10^{-20}~{\rm s\;s^{-1}}\sim
10^{-13}~{\rm s\; yr^{-1}}$. Consequently, their magnetic field $B$, as estimated from
the timing properties, are three to four orders of magnitude weaker than
those of normal radio pulsars, for which $B\simeq 10^{12}$ G.

According to the current theory of neutron star (NS) evolution, millisecond (radio)
pulsars (MSPs) originate in ``radio-dead'' pulsars via the accretion-caused spin-up in
low-mass X-ray binaries (LMXBs, see \citealt{Alpar1982,Radhakrishnan1982}). During this
``recycling'' process, the rotation frequency increases from an initial value $\lesssim 0.1$~Hz to a final
$\sim 500$ Hz in $\sim 10^8-10^9$ yr. The process is associated with the accretion of matter via an accretion disk around the NS. Millisecond
X-ray pulsars become millisecond radio pulsars after the accretion process stops.
This scenario has been corroborated by
the detection of millisecond X-ray pulsations in LMXBs, interpreted as the
manifestation of rotating and accreting NSs \citep {Wijnands1998} and the observations of three objects in transition from a state of accretion with X-ray emission to a rotation-powered state with radio emission and/or vice versa: PSR J1023+0038 \citep{AS09,PA14}, PSR J1824-2452I \citep{PF13} and XSS J12270-4859 \citep{BP14}.

The difference of typically three to four orders of magnitude in the magnetic field strength of MSPs and normal pulsars is explained either by the ``burying'' of the original magnetic field
under a layer of accreted material
\citep{BisnovatyiKomberg1971,Taam1986,Cumming2001} and/or by the Ohmic dissipation
of electric currents in the accretion-heated crust
\citep{Romani1990,GeppertUrpin1994}.

It is expected that the ``recycling'' process in LMXBs is
a particularly widespread mechanism in dense stellar systems such as globular clusters.
This is in accordance with the peculiar structure of the MSP
population \citep{Lorimer2008}: out of a total of about $260$ MSPs,
$\sim60\%$ are in binaries, whereas for other (non-millisecond) pulsars,
this percentage is one order of magnitude lower ($\sim 4$\%).
Simultaneously, about
half (117) of all MSPs are located in $23$ Galactic
globular clusters. Finally, some 50\% of the globular cluster
MSPs are found in binary systems\footnote{ATNF Pulsar
Catalogue\\ {\tt http://www.atnf.csiro.au/people/pulsar/psrcat}} \citep{ATNF}.

The widely accepted recycling mechanism in LMXBs suggests that rapid
MSPs (say those with $P<5~$ms) are likely to be massive.
Therefore, they are important
 for the observational determination of the maximum allowable mass for NSs.
This upper bound is a crucial constraint on the poorly known equation of
state (EOS) at supra-nuclear density. The precise measurement of the mass $M=1.67\pm 0.02~\msun$
of PSR J1903+0327 \citep{Freire11} and, even more so, of $1.97\pm 0.04\ \msun$
for PSR J1614$-$2230 \citep{Demorest2010} confirms that the population of MSPs
contains massive NSs. The most massive pulsar to date is PSR J0348+0432 with $M=2.01\pm 0.04\ \msun$ \citep{Antoniadis2013}. Its properties are a relatively long spin period ($P=39$ ms) and a short orbital period ($P_{\rm orb}=2.46$ hr) for a recycled pulsar. Combined with the low mass of its helium white-dwarf (WD) companion, $M_{\rm WD}=0.17\ \msun$, the case of PSR J0348+0432 is challenging for stellar evolution theory (see e.g. \citealt{Antoniadis2013}), and its formation will not be addressed in the following. On the other side of the mass spectrum, PSR J0751+1807 only has a mass of $1.26\pm0.14~\msun$ (\citealt{Nice2008-1.26}; all measurements are given at $1\sigma$ confidence level). As of today, the masses
of MSPs are therefore bracketed by $1.26\pm0.14\;\msun$ and
$1.97\pm 0.04\ \msun$. We focus on PSR J1614$-$2230 and PSR J0751+1807, which despite possessing extremely different masses, have rather
standard $P$ and $\dot{P}$ for MSPs and are similar to each other. In this paper, by ``progenitor NS'' we denote the NS as it was born in a supernova explosion.

The binary MSPs PSR J1614$-$2230 and PSR J0751+1807
are both located in the galactic disk. That they do not belong to globular
clusters is important for studying stellar evolution.
Indeed, globular clusters are very dense systems composed of thousands of
stars, where frequent stellar interactions may change the
orbital parameters and sometimes even replace the companion star by another star, erasing
the memory of the previous stages of evolution.

Stellar evolution theory for binary stars aims at reconstructing the different stages at the origin of the formation of PSR J1614$-$2230 and PSR J0751+1807 and at explaining their measured masses and other parameters, the masses of their white dwarf companions, and the parameters
of the binary orbits (collected in Table \ref{table:psr}). In the case of PSR J1614$-$2230, this task was
undertaken by \citet{TaurisLK2011} and \citet{LinRappaport2011}.
While we utilize the results of \citet{TaurisLK2011}, we concentrate on a
refined description of the disk-accretion spin-up process that actually produced
the observed object. The evolutionary scenario of the formation of PSR J0751+1807
will combine various elements of available evolutionary scenarios that
led from a wide binary of two main sequence (MS) stars to the present NS+WD binary.
Our main results refer to the progenitor NSs of PSR J1614$-$2230 and
PSR J0751+1807, which turn out to have very different masses. This may be
interesting in the context of the (still incomplete) theory of the formation of NSs in core-collapse supernov\ae.

The article is composed as follows. In Sect.~\ref{sect:scenarios}, the evolutionary scenarios for the formation
of PSR J1614$-$2230 and PSR J0751+1807 binary systems are
presented. Our model for the recycling process based on \cite{Bejger2011} is summarized in Sect.~\ref{sect:spinup},
and the details about the EOS of dense
matter, the model for the magnetic field, and its accretion-induced decay
are given in Sects.~\ref{sect:EOS}, \ref{sect:Bp}, and \ref{sect:decayB},
respectively. Results for PSR J1614$-$2230 and PSR J0751+1807 and bounds on the
properties of the progenitor NS of these two pulsars are presented in Sects.~\ref{sect:bounds-1.97} and \ref{sect:bounds-1.26}. Section~\ref{sect:spin-down} contains discussions of the age of the two MSPs and of their spin frequency just after the spin-up process ended, and conclusions are given in Sect.~\ref{sect:conclusions}. Additionally, 
in the appendix we present a simple approximation of our model, which, from the currently observed properties of a given MSP, enables us to calculate the parameters of its progenitor NS.
Preliminary results of this work were presented at the CompStar 2011 Workshop
in Catania, Italy, May 9-12, 2011, and in \citet{Bejger2013Beijing}.

\begin{table*}
\caption{Measured parameters of the binary pulsars PSR J1614$-$2230 and
PSR J0751+1807 and the masses of their WD companions.
$B_0$ is the canonical value of the magnetic field obtained using
the dipole formula Eq.\;(\ref{eq:bdip}) for an ortogonal rotator and the ``canonical'' NS radius
and the moment of inertia: $R_0=10$~km and $I_0=10^{45}\ {\rm g~cm^2}$.}
\center\begin{tabular}[t]{ccccccccc}
\hline\hline
 PSR &$M_{_{\rm PSR}}$& $P$& $f$ & $\dot{P}$&$P_{\rm orb}$
 &$e$ &$M_{_{\rm WD}}$&$B_0$ \\
        & $(\msun)$  & (ms)   & (Hz) & ($10^{-21}$)& &
        &  $(\msun)$ & ($10^8$ G) \\
 \hline
    J1614$-$2230 & $1.97\pm 0.04$ & 3.15 & 317 & 9.6  &  8.7 d
     &$1.3\times 10^{-6}$ & $0.500\pm 0.006$&1.76  \\
J0751+1807 & $1.26\pm 0.14$
& 3.48 & 287 & 7.8  &  6.3 h
     &$7.1\times 10^{-7}$ & $0.12\pm 0.02$&1.66  \\
 \hline\hline
\label{table:psr}
\end{tabular}
\end{table*}

%
\section{Evolutionary scenarios of formation of two binaries}
\label{sect:scenarios}
\begin{figure}
            \resizebox{\hsize}{!}{\includegraphics[angle=0,clip]{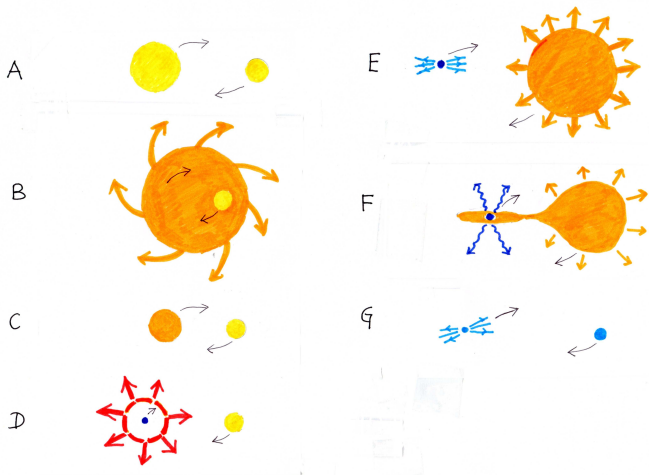}}
            \caption{(Colour online) Main stages of the evolution of the binary system
            leading to the currently observed PSR J1614$-$2230, accompanied by a
             $0.5\ M_\odot$ WD (detailed description in Sect.~\ref{sect:scenario.1.97}).
 {\bf A} - main sequence stage;
 {\bf B} - common envelope stage (secondary inside the primary);
 {\bf C} - the primary becoming a He star;
 {\bf D} - core-collapse supernova explosion of the primary;
 {\bf E} - intermediate-mass X-ray binary stage, with a very strong
 mass loss;
 {\bf F} - LMXB, NS recycling stage via accretion disk, secondary mass
 loss via RLO;
 {\bf G} -  current state, wide binary MSP+WD.}
                \label{fig:evol197}
            \end{figure}

\begin{figure}
            \resizebox{\hsize}{!}{\includegraphics[angle=0,clip]{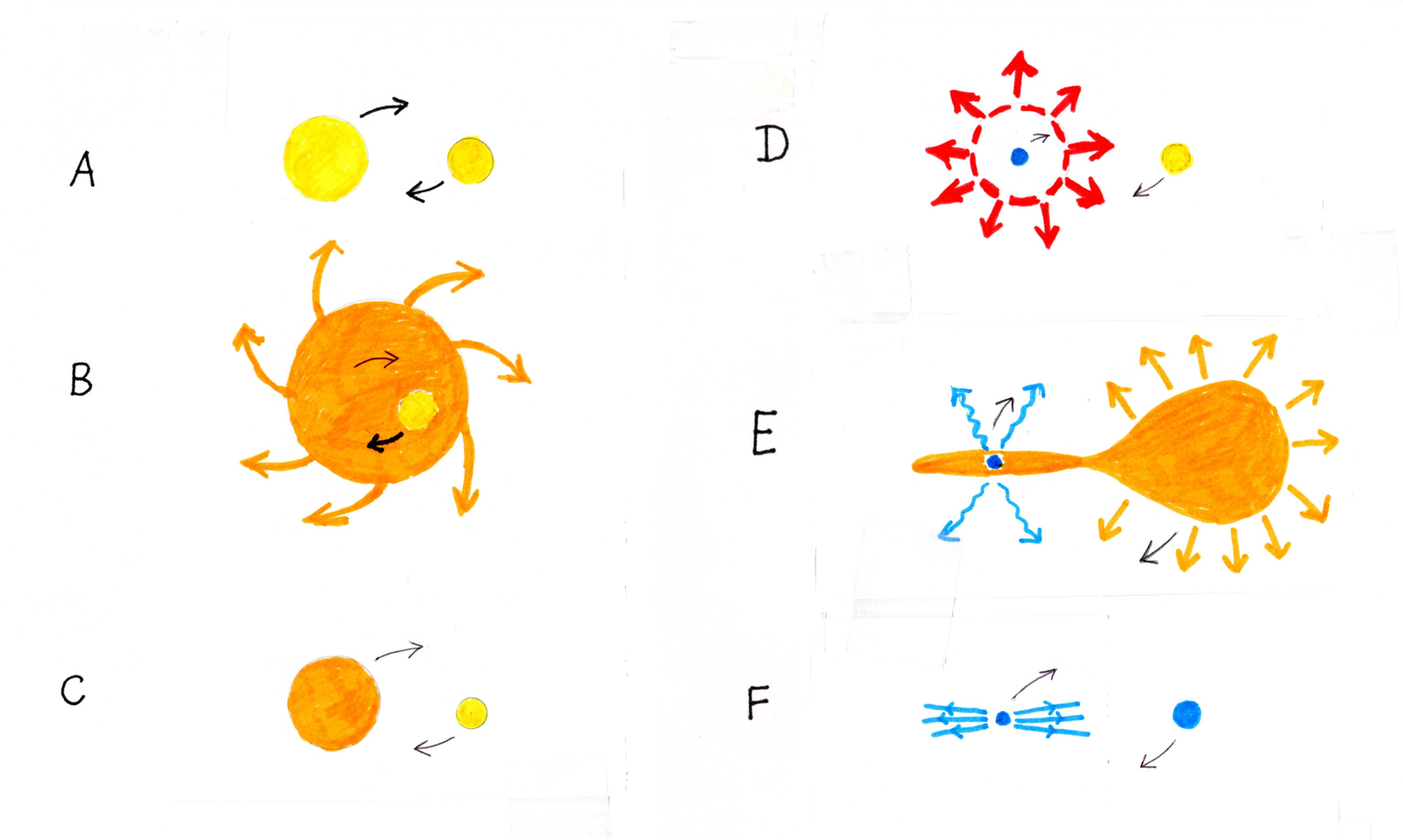}}
            \caption{(Colour online) Main stages of the evolution of the binary system leading to the
            currently observed PSR J0751+1807, accompanied by a $0.12\ M_\odot$ WD
            (detailed description in Sect.~\ref{sect:scenario.1.26}).
 {\bf A} - main sequence stage;
 {\bf B} - common envelope stage (secondary inside the primary);
 {\bf C} - the primary becoming a He star;
 {\bf D} - core-collapse supernova explosion of the primary;
 {\bf E} - Roche lobe filling of the secondary and its strong mass loss,
 LMXB and recycling stage;
 {\bf F} - current state: compact binary MSP+WD.}
                \label{fig:evol126}
            \end{figure}

We begin by sketching the plausible evolutionary scenarios that
could have led to the present binaries containing PSR J1614$-$2230 and
PSR J0751+1807 with their WD companions. We assume that in
both cases, the initial system is a binary of two main sequence stars
of different masses: a more massive primary of initial mass $M_1$
will eventually produce a MSP, while a less massive
secondary of initial mass $M_2$ will become a WD.
The scenarios presented here are but brief and approximate summaries
based on existing work. The main stages of the evolution leading to PSR J1614$-$2230 and PSR J0751+1807
binaries are schematically presented in Figs.~\ref{fig:evol197} and~\ref{fig:evol126}.
For PSR J1614$-$2230 we rely on
\citet{TaurisLK2011} by selecting their Case A, with some
modifications explained in the text, and reviews by
\citet{LooreDoom1993} and \citet{TaurisHeuvel2003}. We use the
evolutionary simulations of \cite{LinRappaport2011} performed in the
case of massive NSs born in SN explosion of the primary star,
presented in Sect.~3.3 of their paper. We are aware that the evolutionary models, particularly those of massive
stars, depend on many assumptions and approximations. Therefore, in
what follows, we restrict ourselves to giving only approximate values of
masses and timescales.

For PSR J0751+1807 binary, for which no detailed evolutionary
calculations exist, we have constructed a scenario using (with some
modifications) the material available in
\citet{LooreDoom1993}, \citet{TaurisHeuvel2003}, and \citet{Istrate2014}.

It has been recently pointed out that the terminal phase of the Roche lobe
overflow (RLO) stage could be associated with loss of angular momentum by the
spun-up NS \citep{Tauris2012Science}. This is related to the expansion of
the magnetosphere accompanied by a decreasing accretion rate and the breaking of the
quasi-equilibrium character of the spin evolution. The gradual switching-off of accretion
occurs during the Roche lobe detachment phase (RLDP). If the duration of this
phase, $t_{\rm _{RLDP}}$, is much shorter than the timescale for transmitting the effect
of braking to the NS $t_{\rm torque}$, then the effect of the RLDP spin-down is
negligible compared to the spin-up during the LMXB stage or the intermediate-mass X-ray binary
 (IMXB) stage. This was shown to be the case for PSR J1614$-$2230 \citep{Tauris2012}.
 This MSP has a CO WD companion and
 was demonstrated to have its IMXB stage
terminated by a rapid ($t_{\rm _{RLDP}}\ll t_{\rm torque}$) RLDP. However, PSR
J0751+1807 has a He WD companion, and it is expected that the LMXB stage terminates
there by a slow RLDP, so that the accreted mass that we calculate when neglecting the
RLDP effect is an underestimate. An example of a slow RLDP is illustrated in Fig.~7 of
\citet{Tauris2012}, but it refers to a binary MSP that is quite
different from PSR J0751+1807. It has $P=5.2$~ms, instead of 3.48 ms for PSR
J0751+1807. Incidentally, in this example the NS
rotation period just before the RLDP coincides with the present period of PSR
J0751+1807 (which we get by construction at the end of our LMXB stage). The expected
effect of including the RLDP braking on the accreted mass required to reproduce
the present period of PSR J0751+1807 will be discussed in
Sect.~\ref{sect:conclusions}.

Predicted masses, timescales, and orbital periods referring to each stage,
collected in Table \ref{tab:197.126.numbers}, are
but approximate estimates. We will stress the differences in the evolutionary
scenarios, conditioned by the present parameters of the pulsars and their white
dwarf companions.

\begin{table*}
\begin{center}
\caption{Summary of evolutionary stages that
led from the two ZAMS binaries to the present millisecond pulsar -
WD binaries, for
 PSR J1614$-$2230 and PSR J0751+1807. The IMXB/LMXB column 
 gives a rough estimate of the duration of the accretion phase onto NS at this stage. The last column contains
 presently measured parameters of the binaries.}
\begin{tabular}[t]{cccccccccc}
\hline\hline
  Pulsar &  & ZAMS  &   1st RG &  CE &   SNIb/c  &  2nd RG  &  IMXB/LMXB &  today \\
 \hline
{PSR J1614$-$2230} & primary  & $25\;\msun$  &  $5\times 10^6$ yr & $7\;\msun$
& $1.9\;\msun$ &    &   &  $1.97\;\msun$ \\
  &secondary  &  $4.5\;\msun$ &   &  $4.5\;\msun$   & $4.5\;\msun$
  &  $5\times 10^7$ yr & $5\times 10^7$ yr & $0.50\;\msun$ \\
  & $P_{\rm orb}$ & $3$ yr  &    &  $4\;$d  &  $2\;$d &    &    &  $8.7\;$d  \\
\hline
{PSR J0751+1807} & primary  & $15\;\msun$  &  $10^7$ yr & $5\;\msun$
& $1.2\;\msun$ &    &   &  $1.26\;\msun$\\
  &secondary  &  $1.6\;\msun$ &   &  $1.6\;\msun$   & $1.6\;\msun$
  &  $3\times 10^9$ yr & $10^9$ yr & $0.12\;\msun$ \\
  & $P_{\rm orb}$ & $5$ yr  &    &  $1\;$d  &  $3\;$d &    &    &  $6.3\;$h  \\
 \hline\hline
\label{tab:197.126.numbers}
\end{tabular}
\end{center}
\end{table*}

\subsection{PSR J1614$-$2230}
\label{sect:scenario.1.97}
Binary parameters and their evolution after the primary SN
explosion are taken from Case A of \citet{TaurisLK2011} and Sect.~3.3
 of \citet{LinRappaport2011}, with some modifications for the sake of
consistency between our scenario for the spin-up of the NS during the LMXB stage
and the present pulsar parameters. In what follows we also use the
reviews of \citet{LooreDoom1993} and \citet{TaurisHeuvel2003}. In view of
the uncertainties in the evolution models, we restrict ourselves to giving only
approximate values of masses and timescales.

\vskip 2mm
 {\bf A: main sequence (MS)} At the zero-age main sequence
(ZAMS), $M_1=25\;\msun$ and $M_2=4.5\;\msun$, and the orbital period
is $\sim 3$ yr. After some $5\times 10^6$ yr
\citep{LooreDoom1993}, the primary becomes a red giant (RG) and its
envelope absorbs the main sequence secondary. The binary
then enters the common envelope stage.

{\bf B and C: common envelope (CE)} The secondary star spirals
within the primary, transferring its angular momentum to a
weakly-bound envelope. As a consequence, the envelope is shed away
on a timescale of $\sim 10^3$ yr. The accretion onto the secondary
within the CE can be neglected. What remains out of the primary is a
helium star with a mass $M_{{\rm He,1}}=7\;\msun$, in a binary with a MS
secondary. For a mass $M_{{\rm He,1}}>8\;\msun$, the star would have
collapsed into a black hole instead of a NS
\citep{TaurisHeuvel2003}. As an outcome of the CE stage, $18\;\msun$
has been ejected from the binary. Frictional dissipation of kinetic 
energy causes the binary to shrink and shortens its orbital period
to $\sim 4$ d.

{\bf D: Supernova (SN)} The outcome of SNIb/c determines the
initial state for the IMXB/LMXB evolution stages
(\citealt{TaurisLK2011,LinRappaport2011}). We assume $M_{\rm
NS}\approx 1.9\;\msun$, higher than NS masses in
\citet{TaurisLK2011} and \citet{LinRappaport2011} in order to be consistent
with our NS spin-up scenario during the LMXB stage. As a result, in our
scenario the helium star of $M_{{\rm He,1}}=7\;\msun$ collapses into
a massive NS of $M_{{\rm NS}}=1.9\;\msun$, while as much as
$5\;\msun$ is ejected in a SNIb/c explosion. The orbit becomes
strongly eccentric. The angular momentum loss during the supernova explosion
 decreases the orbital period to $\sim 2$~d.
 A massive radio pulsar is born in the centre of the supernova. After
a few tens of Myr of magnetic-dipole rotation braking, the pulsar period
increases to a few seconds. While keeping its original surface
magnetic field $\sim 10^{12}~$G, the NS rotates too slowly to
generate radio emission and enters the pulsar graveyard. Then,
${5\times 10^7}$ yr after the ZAMS stage, the secondary leaves the
MS.

{\bf E and F: Intermediate-mass and low-mass X-ray binary} We
adapt the Case A scenario of \cite{TaurisLK2011}, which is consistent with the high
NS mass scenario in Sect.~3.3 of \citealt{LinRappaport2011}, with
some additional comments. The binary enters the stage of
intermediate-mass X-ray binary, called thus because the donor star has an
initial mass of $4.5\;\msun$, which is substantially higher than what is 
characteristic of a donor star in the initial stage of LMXBs 
($<1\;\msun$).
After filling its Roche lobe and starting to lose mass, the
secondary becomes unstable on a thermal timescale of
$\sim 10^6~$yr \citep{Langer2000}. The mass loss via the RLO is $3.4\;\msun$, so 
that the mass of the secondary decreases to $1.1~\msun$ (Fig.~5 of
\citealt{TaurisLK2011}). The mass accreted by the NS is assumed to be
negligible, at most $0.01~\msun$
(\citealt{TaurisLK2011,LinRappaport2011}). 
Then the system
enters the LMXB stage associated with a spin-up
(recycling) of a dead pulsar via an disk accretion from its companion
(donor) star. The RLO is initiated at $P_{\rm orb}=2.2~$d, and the
final period $P_{\rm orb}=9~$d \citep{TaurisLK2011}. 
The
widening of the orbit results from the mass loss from the system, and the magnetic braking is small. 
Accretion onto the NS induces the dissipation of its
magnetic field to its current value $\sim 10^8\;$G inferred from the
measured $P$ and $\dot{P}$. During the $\sim 5\times 10^7$ yr of the
LMXB stage, the NS is spun-up to 317 Hz by accreting matter from the
accretion disk. The NS spin-up is not considered in
\citet{LinRappaport2011}, where only NS mass and $P_{\rm orb}$ are
studied. The LMXB stage ends after the mass loss of the secondary has
stopped, leaving a $0.5\;\msun$ WD. The binary orbit is
circularized during the LMXB stage owing to the tidal dissipation, and
the orbit eccentricity goes down to $e\sim 10^{-6}$. As stressed in
\citet{TaurisLK2011}, the proposed evolutionary scenario ``is only
qualitative''. Both \citet{TaurisLK2011} and \citet{LinRappaport2011}
report that the initial mass of the recycled pulsar is higher than $1.6\pm 0.1\;\msun$, while we
obtain $\sim 1.9\;\msun$ (see Sect.~\ref{sect:bounds-1.97}). As a result of the IMXB/LMXB stage,
dominated by the mass loss from the system with a negligible effect
of the magnetic braking, we obtain a wide binary with $P_{\rm
orb}=8.7\;$d, composed of a WD of $0.5\;\msun$ and a MSP of $1.97\;\msun$ and $f=317\; {\rm Hz}$.

\subsection{PSR J0751+1807}
\label{sect:scenario.1.26}

We assume that during the first 15 Myr, the binary evolution follows 
the one summarized in Fig.~16.12 of \citet{TaurisHeuvel2003}.
We slightly deviate from this evolutionary track by assuming that
the supernova explosion of the primary produces a $1.2\;\msun$ NS.
This mass is $0.1\;\msun$ lower than in \citet{TaurisHeuvel2003}
and should therefore result in a slightly higher eccentricity of the
post-SN binary. We also assume that, because of a weaker angular momentum loss,
 the orbital period after the primary SNIb/c explosion is initially
 3 d (1 d longer than in \citealt{TaurisHeuvel2003}).
Together with the NS mass of $M=1.2\;\msun$,
this is the starting point of an evolutionary track (converging LMXB
in Fig.~2 E) that we select from a large set calculated by
\cite{Istrate2014}. According to these results, magnetic braking
operates at all times between the ZAMS and the final state MSP+WD.
Secondly, wind mass loss from a donor star is much less than the
loss via the RLO mechanism \citep{Istrate2014}. We adjust the spin-up
duration and the mean accretion rate during the LMXB stage to
reproduce the parameters of the present NS+WD binary. All these
changes, which should be taken with a grain of salt, result from the
lack of complete detailed evolutionary calculations for the
currently observed binary with PSR J0751+1807.

 \vskip 2mm {\bf A: Main sequence} At the ZAMS
$M_1=15\;\msun$ and $M_2=1.6\;\msun$, and the orbital period is
$\sim 5$ yr. In ${\sim 10^7}$ yr, the primary becomes a RG and
absorbs the MS secondary, and a brief ($\sim 10^3$ yr)
CE stage follows.

{\bf B and C: Common envelope} The secondary star spirals towards the centre of the
primary, transferring angular momentum to the envelope of the primary.
 As a consequence, the envelope is shed away on a timescale of
$\sim 10^3$~yr. The envelope of $10\;\msun$ is
ejected from the binary, kinetic energy is frictionally dissipated, the orbit shrinks, and
the orbital period is reduced to $1\;$d. The rest of the primary reduces 
 to a helium star with $M_{{\rm
He,1}}=5\;\msun$ in a binary with a MS secondary.

{\bf D: Primary supernova explosion and Roche-lobe overflow by the
secondary} The evolved core of the helium-star primary collapses
into a light NS: $M_{_{\rm NS}}=1.2\;\msun$, with most of the mass
of the primary being ejected in a SNIb/c explosion.
A low-mass
radio pulsar is born at the SNIb/c centre, and the orbital period
increases to 3 d (by construction, 1 d longer than in
\citealt{TaurisHeuvel2003}. This can be easily obtained by tuning
the orbital angular momentum loss via mass loss in SNIb/c used up to
this point). After a few tens of Myr, the NS enters the pulsar graveyard.
 Then, $3\times10^9$~yr after the ZAMS stage,\footnote{ We used $t_{\rm MS}\propto M^{-3}$ for low-mass
stars as deduced from Table 11.2 of \citet{LooreDoom1993}}
the secondary fills its Roche lobe. This is because of a
rapid orbital angular momentum loss associated with a very efficient
magnetic braking. The orbital period shortens by a factor of three
 \citep{Istrate2014}. Then the secondary overflows its Roche lobe,
and the binary enters the stage of the LMXB.

{\bf E: Low-mass X-ray binary} In the following paragraph 
we rely on the modelling of \citet{Istrate2014} for the evolution of 
the LMXB. During this phase. which lasts a few $10^9$ yr,
the pulsar is spun-up by matter falling from an accretion disk; however, we
 estimate that periods of intense accretion, during which the essential of the spin-up takes place,
 occur on a time scale of $10^9$ yr. We adopt this value in the following.
 The NS magnetic field is buried
by the accreted matter, decreasing to a value $\sim 10^8\;$G derived from the
present $\dot{P}$. The value of $P_{\rm orb}=3$~d at the beginning
of mass transfer is below the bifurcation period $P_{\rm bif}\sim
4\;$d, in contrast to Fig.~16.12 of
\citet{TaurisHeuvel2003}. The LMXB track is therefore of converging
type, and $P_{\rm orb}$ decreases in time (see
\citealt{TaurisHeuvel2003} and references therein). The orbital
period shortens because of an efficient angular momentum loss resulting
from magnetic braking and gravitational wave radiation. 
In what follows we illustrate our case
using an evolutionary track from a large set of tracks calculated by
\citet{Istrate2014}. Taking the LMXB model with initial
$P_{\rm orb}=3.2~$d and adjusted magnetic braking, one gets a compact NS+WD binary with $P_{\rm orb}=6.3~$~h. The NS is
spun-up to 287 Hz by accretion from the disk, and the WD mass at the end
is $0.16\;\msun$, which is quite close to the measured mass of WD.
The efficiency of accretion
 onto the NS is rather low (30\%), so that 70\% of mass lost by
 the secondary leaves the binary. We assume that
 an appropriate small tuning of the LMXB stage can result in a decrease in the WD mass to a measured
 value of $0.12\;\msun$. Finally, after the accretion onto the NS
 has stopped, the pulsar activity restarts.

{\bf F: Current state} The evolution of the binary at the LMXB
stage is dominated by the angular momentum loss associated with an
efficient magnetic braking. As a result, the present MSP+WD
binary is relativistic with an orbital
period of only 6.3 h. The orbit, which is highly eccentric after the SNIb/c
explosion, has been circularized by the tidal interactions of the NS
with the secondary down to $e\sim 10^{-6}$.

\subsection{Differences in evolutionary scenarios and their causes }
\label{sect:scen-differences}
We have assumed that both binary systems, hereafter
referred to as H (heavy) and L (light), resulted from the evolution of
binaries that originally consisted of two ZAMS stars. Therefore, the striking
differences between today's NS+WD binaries are the imprint of initial
conditions. NS(H) originated in a $25\;\msun$ ZAMS primary star, to be
compared with a $15\;\msun$ ZAMS progenitor for NS(L). Even more dramatic is the
difference between the ZAMS masses of progenitors of WDs:
$M_2({\rm H})=4.5\;\msun\approx 3M_2({\rm L})$.

The ZAMS masses of stars in the H binary are significantly higher than the masses
of their counterparts in the L binary. Consequently, H-binary evolution timescales
are significantly shorter than L-binary ones. The formation of the first RG star in the
H binary requires half of the time needed for this in the L binary. The second RG in
the H binary is formed after $5\times 10^7$ yr, which is only $\sim2\%$ of the time
needed for that in the L binary.

As much as $18\;\msun$ is lost by the H binary during the CE stage, nearly double the mass loss by the L binary during CE evolution.

The H-binary orbital evolution during the IMXB-LMXB stage is
dominated by the mass loss, the effect of magnetic braking being absent, and
$P_{\rm orb}$ increases from 2~d after SNIb/c explosion to
the current value of 8.7~d. The L binary only goes through the LMXB stage. In
contrast to the H case, L-binary orbital evolution has to be
dominated by the angular momentum loss caused by the magnetic
braking, and this allows $P_{\rm orb}$ to decrease from 3 d just
after the SNIb/c, down to 6 h today, so that $P_{\rm orb}({\rm
H})=35P_{\rm orb}({\rm L})$.
\section{Spin-up by accretion}
\label{sect:spinup}

We now briefly summarize the model for the recycling process, i.e. the
spin-up of a progenitor NS by accretion of matter from a thin disk leading to
the formation of a millisecond pulsar. We do not model the evolution of the binary system
consisting of the NS and its companion, but only the spin-up of the NS due to accretion of matter.
Our approach follows \citet{Bejger2011}
and is applied here to the two pulsars PSR J1614$-$2230 and PSR J0751+1807. We assume that the
evolution of an accreting NS can be described by a sequence of stationary
rotating configurations of increasing baryon mass $\mbar$, obtained for an assumed EOS.
The increase in total stellar angular momentum $J$ is calculated by taking
into account the transfer of specific orbital angular momentum of a
particle accreted from a thin accretion disk. This proceeds at a distance $r_0$ from the
centre of the NS. It results from the interaction of the disk with the NS magnetic field and is obtained using the
prescription for the magnetic torque by \citet{klurap}.

The spin evolution of an accreting NS results from the interplay between the
spin up due to the accretion of matter associated with angular momentum
transfer and the braking due to the interaction between the NS magnetic field and the
accretion disk. For details of the implementation 
and tests, we refer to \citet{Bejger2011}. The evolution of the
total angular momentum transferred to the NS of mass $M$ and radius $R$ is described by 
\begin{eqnarray}
 \frac{{\rm d}J}{{\rm d}\mbar}=\ltot&\equiv& l(r_0) -\lmag \nonumber\\
&=&l(r_0) -\frac{\mu^2}{9r_0^3 {\dot M}}
 \left(3-2\sqrt{\frac{r_c^3}{r_0^3}}\right),
 \label{eq:ltot}
\end{eqnarray}
 with the magnetic moment $\mu=BR^3$. Here, $l(r_0)$ is the specific angular momentum of
an infalling particle at the inner boundary of the disk $r_0$
and $\lmag$ (magnetic torque divided by the accretion rate\footnote{ For simplicity we adopt the notation $\md$ for the accretion rate instead of $\dot{M}_{\rm B}$. The accretion rate is indeed described in terms of baryon mass $\mbar$, since this number is well defined
for the binary system, see \cite{Bejger2011}.}) describes the interaction
of the NS magnetic field with the accretion disk. One can define three characteristic lengths of the problem: the magnetospheric radius $\rmag=\left(GM\right)^{-1/7}\dot{M}^{-2/7}\mu^{4/7}$,
 the corotation radius $\rcor=\left[GM/(4\pi^2f^2)\right]^{1/3}$, and the location of the relativistic
marginally stable orbit $\rms$. Then, the inner boundary
$r_0$ is determined by the condition of vanishing
viscous torque leading to an algebraic equation containing these lengths:
\begin{equation}
\label{bc}
\frac{1}{2} f_{\rm ms}(r_0) = \left(\frac{r_m}{r_0}\right)^{7/2} \!\! \left(\sqrt{\frac{r_c^3}{r_0^3}}-1\right)\, {\rm with }\,
f_{\rm ms}(r)=\frac{2}{\Omega r} \frac{{\rm d}l}{{\rm d}r}
,\end{equation}
which is a dimensionless function describing the location of the marginally stable orbit for rotating NSs in General Relativity, defined by
$f_{\rm ms}(\rms)=0$.

The modeling of the accretion phase consists in adjusting the value of the mass $M_0$
 (or equivalently the baryonic mass $M_{{\rm B}0}$), the magnetic field $B_{\rm i}$ of the progenitor NS, and the
mean accretion rate $\dot{M}$ (accretion rate averaged over the whole accretion process) so that at the end of the accretion process,
the NS has its parameters, i.e. its mass $M$ (baryonic mass $M_{\rm B}$) and magnetic field $B$, coinciding with those of a given millisecond pulsar. We also consider that the post-accretion frequency is equal to its currently observed value $f$. The validity of this assumption is discussed in Sect.~\ref{sect:spin-down}.
For a given final configuration (i.e. $M$, $P$, and $B$), a family of sets of three parameters: $M_0$
(hence the amount of accreted matter $M_{\rm acc}=M_{\rm B}-M_{{\rm B}0}$), $B_{\rm i}$ and $\dot{M}$
(or equivalently the duration of the accretion phase $\tau_{\rm acc}=M_{\rm acc}/\dot{M}$), is obtained.
As a consequence, a choice of $\tau_{\rm acc}$ imposes $M_{\rm acc}$ and $B_{\rm i}$.


\section{Equations of state}
\label{sect:EOS}
The EOS of the dense cores of NSs is still poorly known. This is due to,
on the one hand, a lack of knowledge of strong interaction in dense matter and, on the
other hand, deficiencies in the available many-body theories of dense matter. This
uncertainty is reflected in a rather broad scatter of theoretically derived and EOS-dependent maximum allowable mass for NSs (see e.g.
\citealt{NSbook2007}). Fortunately, the recent measurements of the mass of PSR
J1614$-$2230 as $M=1.97\pm 0.04~{\rm M}_\odot$ \citep{Demorest2010} and PSR J0348+0432 as $M=2.01\pm 0.04~{\rm M}_\odot$ \citep{Antoniadis2013} introduce a fairly
strong constraint on $M_{\rm max}$. It implies that the (true) EOS is rather
stiff. To illustrate the remaining uncertainty, we considered two
different models for dense matter:


\begin{itemize}

\item[] {\bf DH} \citep{DouchinH2001}. It is a non-relativistic model for the simplest possible composition of matter: neutrons, protons, electrons, and muons in $\beta$ equilibrium.
The energy density functional is based on the SLy4 effective nuclear interaction. The
model describes both the dense liquid core of the NS and its crust in a unified way.
This EOS yields a maximum mass $M_{\rm max}=2.05~\msun$ and a circumferential radius at maximum mass
$R_{M_{\rm max}}=10.0~$km (for a non-rotating configuration).
\vskip 2mm
\item[] {\bf BM} \citep{Bednarek2011}. It is a nonlinear relativistic model that allows for a softening owing to the appearance of hyperons at a density of $\sim
6\times 10^{14}~{\rm g~cm^{-3}}$. The nonlinear Lagrangian includes up to quartic terms in the meson fields. The meson fields $\sigma,\omega$, and $\rho$ are coupled to nucleons and hyperons, and hidden-strangeness meson fields
$\sigma^\star$ and $\phi$ only couple to hyperons. The vector meson $\phi$ generates
high-density repulsion between hyperons. The EOS is calculated in the mean field
approximation. It yields $M_{\rm max}=2.03~\msun$ and $R_{M_{\rm max}}=10.7~$km (for a non-rotating configuration).
\end{itemize}

Our models for baryonic matter do not include the $\Delta$ resonance (first
excited state of the nucleon) as a real constituent of dense NS matter. In 
the vacuum, the $\Delta$ is
 very different from hyperons. The instability of hyperons is due to
the weak interactions, with a typical half width to the rest-energy ratio
for the lightest hyperon $\Lambda$: $\Gamma_\Lambda/m_\Lambda
c^2\approx 10^{-14}$. Hyperons in the vacuum are stable with respect to
the strong interaction processes. In contrast, the $\Delta$ is unstable because of
the strong interactions, and $\Gamma_\Delta/m_\Delta c^2\approx 0.1$. In
dense matter, the rest energy of a baryon should be replaced by the
self-energy (in-medium mass), when taking the interactions with
other particles into account, as well as the exclusion principle blocking some
processes that were allowed in the vacuum. For example, when the $\Delta$ is
stabilized in dense matter, this means a decrease in $\Gamma_\Delta$
 from 120 MeV in the vacuum to zero in dense matter.
 It has been argued that these medium-induced effects increase the threshold density for the
appearance of stable $\Delta$ in dense matter above the maximum density
reachable in NSs (e.g. \citealt{Sawyer1972, Nandy1974,
Glendenning1985}). In the relativistic mean field approximation, the
threshold density for the $\Delta$ depends strongly on the in-medium $\Delta$
mass and the completely unknown $\Delta$ coupling constant to the isovector-vector 
meson $\rho$. Recently, the question of a possible presence of $\Delta$ in NSs was
revived in \citet{Drago2014a,Drago2014b}. Viewing the uncertainties in
the in-medium effects on the $\Delta$ at supra-nuclear densities, however,
we do not consider them in NS cores.

\section{Magnetic field of a pulsar}
\label{sect:Bp}
Only an estimate of the magnetic field $B$ of a given pulsar can be obtained if its rotational period $P$ and period derivative $\dot{P}$ are known.
Assuming that the pulsar is a rotating magnetic dipole and that its loss
of rotational energy originates in the emission of magneto-dipole
radiation alone, one derives the following classical dipole formula (see
e.g. \citealt{NSbook2007}):
\begin{equation}
B=\sqrt{\frac{3c^3IP\dot{P}}{8\pi^2 R^6} \frac{1}{\sin^2 \alpha}},
\label{eq:bdip}
\end{equation}
where $\alpha$ is the angle between the rotation and magnetic axes: $0<\alpha\leqslant 90^\circ$.
This formula describes a spinning dipole in a vacuum, meaning a dipole without plasma in the magnetosphere. It breaks down
for the case of aligned rotator, $\alpha=0^\circ$.
However, a more physically sound formula has been derived by \citet{Spitkovsky2006}, approximating solutions of the force-free relativistic MHD equations in the magnetosphere filled with plasma for both aligned and oblique rotators. For $0\leqslant \alpha \leqslant90^\circ$,
\begin{equation}
B=\sqrt{\frac{c^3IP\dot{P}}{4\pi^2 R^6} \frac{1}{1+\sin^2 \alpha}}.
\label{eq:bspi}
\end{equation}

To test the dependence of the modelling on the estimate of the magnetic field of the pulsar during the recycling process, two different models are considered in the following, each of them corresponding to a given value of the parameter $\beta$ in the equation
\begin{equation}
B=\beta \sqrt{\pms\pd_{-20}}\frac{\sqrt{I_{45}}}{R^{3}_{6}}\,10^8\, \textrm{ G}.
\label{eqn:bp}
\end{equation}
In the above formula, $\pms$ is the period in ms, $\pd_{-20}$ the period derivative in units of $10^{-20}$~s~s$^{-1}$, $I_{45}$ the moment of inertia in units of $10^{45}$ g cm$^2$, and $R_6$ the radius in $10^6$ cm. The dimensionless parameter $\beta$ takes the following value:
\begin{enumerate}
\item[] {\bf Model (a)}: $\beta=1.01$. This corresponds to taking the lowest possible value of the magnetic field for the magnetic dipole obtained from the dipole formula (\ref{eq:bdip}) for an orthogonal rotator ($\alpha=90^\circ$);
\item[] {\bf Model (b)}: $\beta=0.83$ derived using Spitkovsky's formula (\ref{eq:bspi}) for an aligned rotator ($\alpha=0^\circ$).
\end{enumerate}
The second model is consistent with the model of an accretion disk
presented in \citet{klurap}, as used in our approach that assumes that the magnetic field and the rotation axes are aligned. The first model, which is widely used in the literature, serves as a reference for comparison with previous works (e.g. \citealt{Bejger2011}).

With the above formula, one can calculate the value of the magnetic field of the pulsar at the end of the spin-up process consistently with the EOS, by taking the values of $I$ and $R$ for a NS rotating at the frequency and mass of interest.
In the last column of Table \ref{table:psr}, values $B_0$ of the canonical derived magnetic field are quoted, i.e., values obtained from Eq.\;(\ref{eqn:bp}) for the magnetic field model (a) and considering a NS with canonical values for the radius and the moment of inertia: $R_0=10$~km and $I_0=10^{45}\ {\rm g~cm^2}$.

\section{Decay of NS magnetic field in LMXBs and IMXBs}
\label{sect:decayB}
Although observations do not give any evidence of magnetic field decay
during the radio pulsar phase, a substantial magnetic field
decay (by some four orders of magnitude) is expected to occur during
the ``recycling'' process in a LMXB, leading to the formation
of a millisecond pulsar (\citealt{Taam1986}, for a review see
\citealt {Colpi2001}).

The theoretical modelling of the accretion-induced decay of $B$ is a
challenging task (\citealt{ZK06}, for a review see \citealt{Bejger2011}). Considering current uncertainties in the modeling of the magnetic field
decay accompanying the spin-up phase of a given
millisecond pulsar, we employ phenomenological
models, based to some extent on observations
of NSs in LMXBs. \citet{Taam1986} analysed several
LMXBs of different ages. They suggested a possible inverse correlation
between $B$ and the total amount of accreted material. This
suggestion was later confirmed in a study by \citet{vdHeuvel1995}.
After analysing a subset of LMXBs, \citet{Shibazaki1989} proposed
to approximate the inverse correlation between $B$ and accreted mass
by a formula:
\begin{equation}
B(M_{\rm acc})=
B_{\rm i}/(1 + M_{\rm acc}/ m_{\rm B}).
\label{eq:B.DeltaM}
\end{equation}
The scaling constant $m_{\rm B}$
controls the pace of dissipation of $B$ with
increasing $M_{\rm acc}$. The values $m_{\rm B}= 10^{-4}$, $10^{-5}~$M$_\odot$ are both consistent with the observed or estimated $P,B,$ and $M_{\rm acc}$ of binary and isolated millisecond radio pulsars \citep{Shibazaki1989,Fran}. In the following, we adopt the value $m_{\rm B}= 10^{-5}~$M$_\odot$ unless stated otherwise.
The limitations and uncertainties of Eq.\;(\ref{eq:B.DeltaM}) have been reviewed
in detail in \citet{Bejger2011}. We use Eq.\;(\ref{eq:B.DeltaM}) as our
baseline description for magnetic field decay in LMXBs and IMXBs. However, we also tested other phenomenological models for the magnetic field decay, such as the ones introduced in \citet{Kiel08}, \citet{Oslowski11}, and \citet{Wijers97}. Our results depend weakly on the choice of a specific model (for details see \citealt{MorganePhD}).

The aim of our analysis is to obtain evolutionary tracks that lead to the formation of a millisecond pulsar whose mass, rotational frequency, and magnetic field are equal to the currently observed values of PSR J1614$-$2230 or PSR J0751+1807 in order to infer some properties of the progenitor NS of these two extreme millisecond pulsars.

\section{PSR J1614$-$2230: lower bounds on $\dot{M}$ and the progenitor NS mass}
\label{sect:bounds-1.97}
\begin{table}
\begin{center}
\caption{PSR J1614$-$2230: Values for a NS rotating at $f=317$~Hz of the moment of inertia $I_{45}$ (in units of $10^{45}$ g cm$^2$), the equatorial radius $R_6$ (in $10^6$ cm), and of the magnetic field $B_8$ (in $10^8$ G). The last is obtained from Eq.\;(\ref{eqn:bp}) for the different models of magnetic field, the two EOSs, and the extrema and central values of the measured mass interval at $1\sigma$ level.}
\begin{tabular}{cccccc}
\hline \hline
EOS & Mass ($\msun$) & $B$ model & $I_{45}$ & $R_6$ & $B_8$ \\
\hline
DH & 1.97 & (a) & 1.99 & 1.08 & 1.98 \\
DH & 1.97 & (b) & 1.99 & 1.08 & 1.62 \\
BM & 1.97 & (a) & 2.19 & 1.18 & 1.57 \\
BM & 1.93 & (a) & 2.22 & 1.21 & 1.48 \\
BM & 2.01 & (a) & 2.10 & 1.13 & 1.77 \\
\hline
\hline
\label{tab:bmax}
\end{tabular}
\end{center}
\end{table}

\begin{figure}
            \resizebox{\hsize}{!}{\includegraphics[]{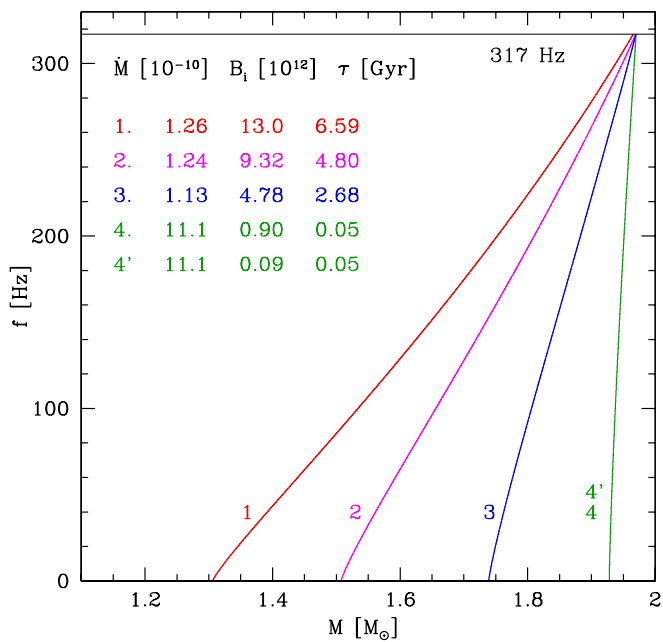}}
            \caption{(Colour online) Example of spin-up tracks of accreting NSs leading to the final configuration of PSR J1614$-$2230 for the BM EOS and for model (a) for the magnetic field decay. For each track, the mean accretion rate (in $\msolyr$), the
            initial magnetic field (in G), and the duration of the accretion phase are indicated.}
            \label{fig:fmhigh}
\end{figure}

\begin{figure}
            \resizebox{\hsize}{!}{\includegraphics[]{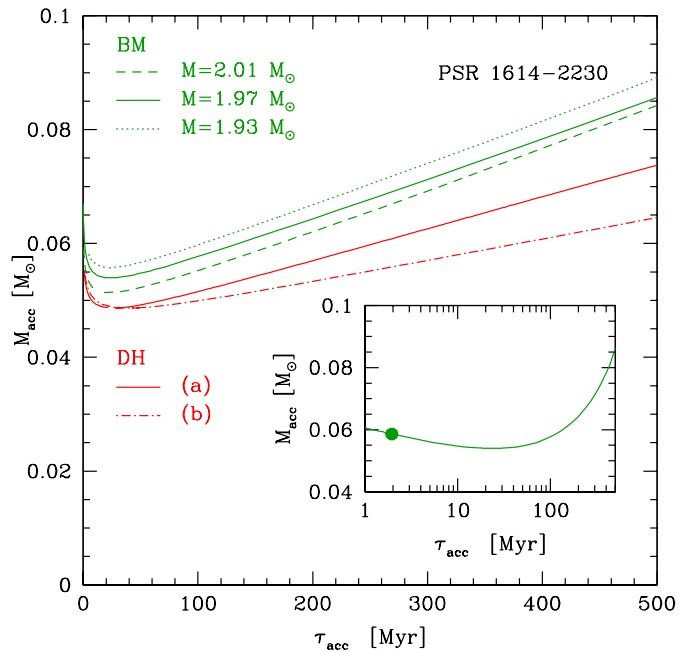}}
            \caption{(Colour online) Mass of accreted matter as a function of the time needed to spin-up
            the progenitor NS to the observed properties of PSR J1614$-$2230 (i.e., $M$, $B,$ and $f$)
            for the five models indicated in Table \ref{tab:bmax}. The dot corresponds to a spin-up track proceeding at a rate equal to the Eddington
            accretion rate $\medd \sim 3\times 10^{-8}\msolyr$; the accretion timescale is then $\sim 2$~Myr. For tracks with $\tau_{\rm acc} \lesssim 2$ Myr, $\dot{M}$ is greater than $\medd$.}
            \label{fig:tmacchigh}
\end{figure}

\begin{figure}
            \resizebox{\hsize}{!}{\includegraphics[]{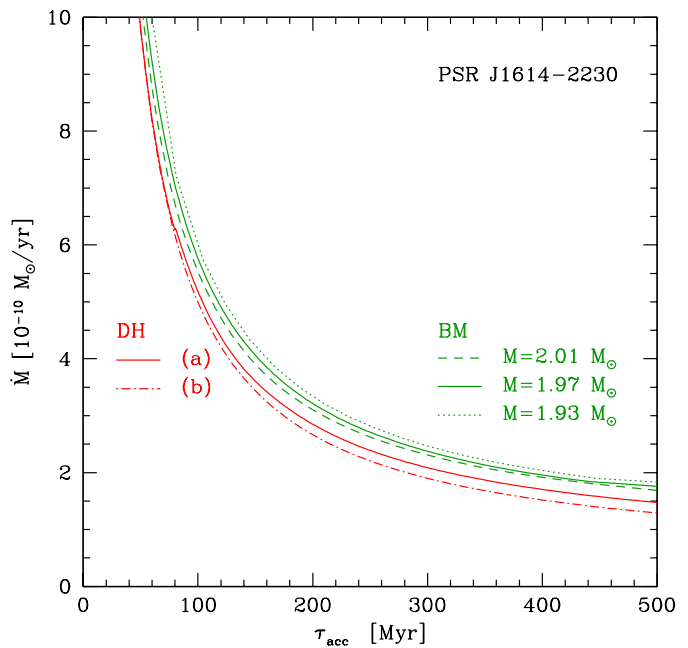}}
            \caption{(Colour online) Mean accretion rate
            versus duration of the accretion phase for DH and BM EOS needed to reach PSR J1614$-$2230 configuration.}
            \label{fig:mdotthigh}
\end{figure}

Five different setups
are employed in modelling the accretion phase leading to the PSR J1614$-$2230 configuration: for the DH EOS, the two models for the magnetic field
are used, and for the BM EOS, three values of the mass of the millisecond
pulsar corresponding to the lower, upper, and central values of the $1\sigma$ mass
interval are taken into account. Their properties are indicated in Table \ref{tab:bmax}.

In Fig.~\ref{fig:fmhigh} different spin-up tracks, i.e. the change in the spin frequency with the mass
of the accreting NS, are shown for the BM EOS, the model (a) for the magnetic field, and for a set of initial (pre-accretion) masses, magnetic fields of the progenitor
NS, and accretion rates. At the end of the recycling process, the final
$M$, $B,$ and $f$ match the present-day parameters of PSR J1614$-$2230.

As noted in \citet{Bejger2011}, spin-up tracks with a non-zero initial frequency, which is expected for newly born NSs, are indistinguishable from those with a zero initial frequency on a time scale of $10^6$ years, which is much shorter than the ones typical of the recycling process. Spin-up tracks 1, 2, and 3 in Fig.~\ref{fig:fmhigh} are only plotted for illustrative purposes since the duration of the accretion phase needed to reach PSR J1614$-$2230 configuration is too long to be physically relevant in accordance with Sect.~\ref{sect:scenarios}.

Tracks 4 and $4^\prime$, which are indistinguishable from one another, are obtained for $m_{\rm B}=10^{-5}$ and $10^{-4}~$M$_\odot$, respectively.
For a given set of final magnetic field, mass, and frequency, models with
$m_{\rm B}=10^{-4}$ and $10^{-5}~$M$_\odot$ give equal values for the
initial mass and the accretion rate. As a consequence of equation
(\ref{eq:B.DeltaM}), initial magnetic fields are one order of magnitude
larger for tracks with $m_{\rm B}=10^{-5}~$M$_\odot$ than for the ones
with $m_{\rm B}=10^{-4}~$M$_\odot$. Track 4, calculated for $m_{\rm
B}=10^{-5}~$M$_\odot$, is obtained for an initial magnetic field $B_{\rm
i}\sim10^{12}~$G, consistent with the inferred magnetic field of
isolated radio pulsars $B\sim10^{11}-10^{13}~$G \citep{ATNF}. For the
value $m_{\rm B}=10^{-4}~$M$_\odot$ used in \citet{Bejger2011,Bejger2013Beijing} and shown by track $4^\prime$, $B_{\rm i}\sim10^{11}~$G.

In the following two quantities are used: the amount of accreted matter given by the relation $M_{\rm acc}=M_{\rm B}-M_{{\rm B}0}$ and the duration of the accretion phase: $\tau_{\rm acc}=M_{\rm acc}/{\dot M}$.
To reach the mass of PSR J1614$-$2230, a low-mass progenitor NS has to accrete more matter than a high-mass one so it undergoes a stronger decay of its magnetic field according to Eq.\;(\ref{eq:B.DeltaM}).
Consequently, its pre-accretion magnetic field had to be larger. The duration of the accretion phase is also longer for a low-mass progenitor than a high-mass one.
Figures~\ref{fig:tmacchigh} and \ref{fig:mdotthigh} show the relation between the accretion time $\tau_{\rm acc}$ and the amount of accreted matter $M_{\rm acc}$ and the mean accretion rate needed to reach PSR J1614$-$2230 in its current configuration. Both figures show weak dependence on the
final NS mass and on the model of magnetic field (i.e. magnetic dipole or Spitkovsky's model). Results for $m_{\rm B}=10^{-4}$ and $10^{-5}~$M$_\odot$ are indistinguishable.

The minimum amount of accreted mass necessary for a NS to become a 
millisecond pulsar is reached for a finite value of $\tau_{\rm acc}$, which corresponds to a 
minimum of the function $M_{\rm acc}(\tau_{\rm acc})$ (see Figs.~\ref{fig:tmacchigh} and \ref{fig:tmacclow}).
 For a final configuration corresponding to PSR J1614-2230, the minimum is obtained for
$\dot M^{\rm min} \sim 2\times 10^{-9}\;\msolyr$
and $\tau_{\rm acc}^{\rm min} \sim 20 - 25$ Myr. 
The existence of this minimum is a consequence of the fact that for accretion rates higher than $\dot M^{\rm min}$, the magnetic torque,
or equivalently $l_{\rm m}$ in Eq.~(\ref{eq:ltot}), changes its sign, becoming positive (see Appendix). The value of $\tau_{\rm acc}^{\rm min}$ is significantly
less than the timescales presented in Figs.~\ref{fig:fmhigh} and \ref{fig:fmlow},
 and along all the tracks discussed in the following, $l_{\rm m}$ is always positive and thus counteracts $l(r_0)$. 
 For an unrealistically short duration of the recycling $\tau_{\rm acc}\lesssim 2$~Myr, the model
 predicts that accretion proceeds at a higher rate than the Eddington rate $\medd \sim 3\times 10^{-8}\;\msolyr$
 \citep{Tauris2012}, as indicated in Fig.~\ref{fig:tmacchigh}.

Model (b) gives lower values for $B$ than dipole model (a) at all stages of the recycling process. Therefore, the magnetic torque that opposes the spin up of the accreting NS is less for model (b), making the recycling process more efficient. Thus less accreted matter is needed to reach the current pulsar parameters.
 Moreover, since $M_{\rm acc}$ is lower for model (b), so is the mean accretion rate in the recycling process for given accretion time $\tau_{\rm acc}$ (see Fig.~\ref{fig:mdotthigh}).

Using evolutionary arguments one can constrain the minimal birth mass $M_0$, at the onset of accretion. For an accretion phase lasting at most $\simeq 50~$Myr (see Sect.~\ref{tab:197.126.numbers}
 and \citealt{TaurisLK2011}), the progenitor NS must have accreted less than $\simeq 0.06\ \msol$. Therefore the progenitor NS was born as massive independently of the EOS: $M_0\simeq 1.9\ \msol$. Such a configuration is illustrated by the spin-up track 4 in Fig.~\ref{fig:fmhigh}.
The value $M_{\rm acc}\simeq 0.06\ \msol$ should be considered as a lower limit, since we do not model the evolution of the binary system as in \cite{TaurisLK2011} or take the possible ejection of matter from the magnetosphere or instabilities in the accretion disk into account. Moreover, in our model the spin-up is assumed to be maximally efficient i.e., all angular momentum from the accreted matter is transferred to the NS. If the efficiency of the accretion process is reduced by 50 per cent (see discussion in \citealt{Bejger2011}), then our calculations show that the accretion of $M_{\rm acc}\simeq0.11~\msun$ is necessary to spin up the NS to PSR J1614$-$2230 current configuration in 50 Myr. The same amount of accreted matter is required if accretion does not proceed from $r_0$ but from the magnetospheric radius $r_{\rm m}$ and with no magnetic torque, unlike in Eq.\;(\ref{eq:ltot}). Finally, as shown in \citet{Bejger2011}, if accretion proceeds from the marginally stable orbit $r_{\rm ms}$, $M_{\rm acc}=0.076~\msun$. Therefore the mass $M_0\simeq 1.9\ \msol$ of the NS at birth is an upper limit.

Assuming that the accretion time is well-constrained, one can also estimate
the mean accretion rate using the currently observed
parameters of PSR J1614$-$2230. Figure~\ref{fig:mdotthigh} gives $\dot{M}>
\left(1.0-1.4\right)\times 10^{-9}\ \msolyr$, the lower value corresponding to the DH EOS.

Modelling the evolution and dynamics of the binary system, \citet{TaurisLK2011} calculated that the NS accreted $\sim 0.3~\msol$ during 50 Myr, i.e. an averaged accretion rate of $\dot{M}\simeq6\times 10^{-9}\ \msol$/yr. The discrepancy between these results and ours stems from the fact that we do not model the evolution of the binary and of the donor star and, to a lesser extent, from a different model for the spin-up phase.

%

\section{PSR J0751+1807: lower bounds on $\dot{M}$ and $M_0$}
\label{sect:bounds-1.26}
\begin{table}
\caption{PSR J0751+1807: Analogue of Table \ref{tab:bmax}.}
\begin{tabular}{cccccc}
\hline \hline
EOS & Mass ($\msun$) & $B$ model & $I_{45}$ & $R_6$ & $B_8$ \\
\hline
DH & 1.26 & (a) & 1.19 & 1.19 & 1.09 \\
DH & 1.26 & (b) & 1.19 & 1.19 & 0.88 \\
BM & 1.26 & (a) & 1.51 & 1.36 & 0.80 \\
BM & 1.12 & (a) & 1.29 & 1.37 & 0.73 \\
BM & 1.40 & (a) & 1.74 & 1.36 & 0.87 \\
\hline
\hline
\label{tab:bmin}
\end{tabular}
\end{table}

\begin{figure}
            \resizebox{\hsize}{!}{\includegraphics[]{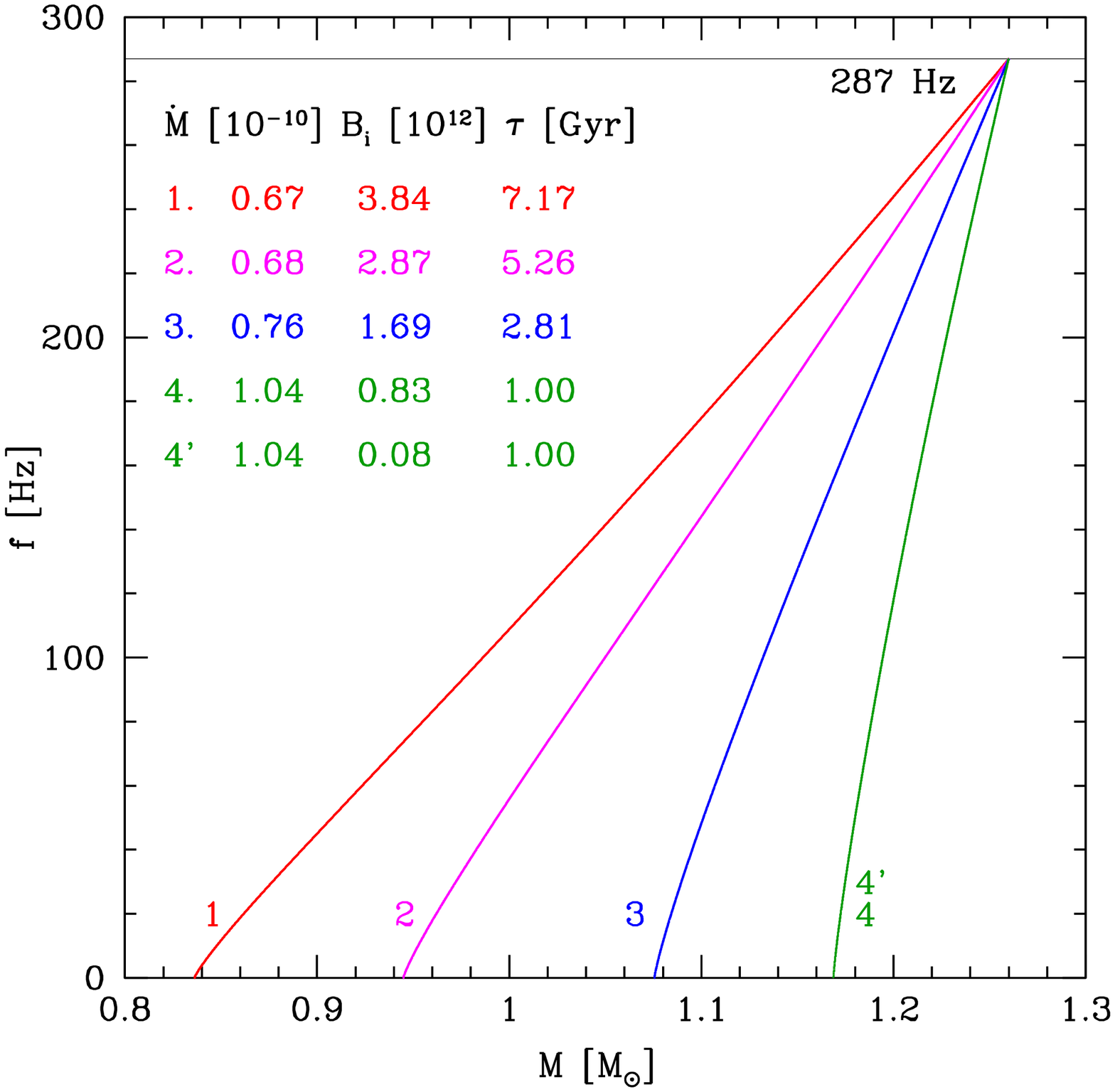}}
            \caption{(Colour online) Similar to Fig.~\ref{fig:fmhigh}, but for the pulsar J0751+1807.}
            \label{fig:fmlow}
\end{figure}

\begin{figure}
            \resizebox{\hsize}{!}{\includegraphics[scale=1]{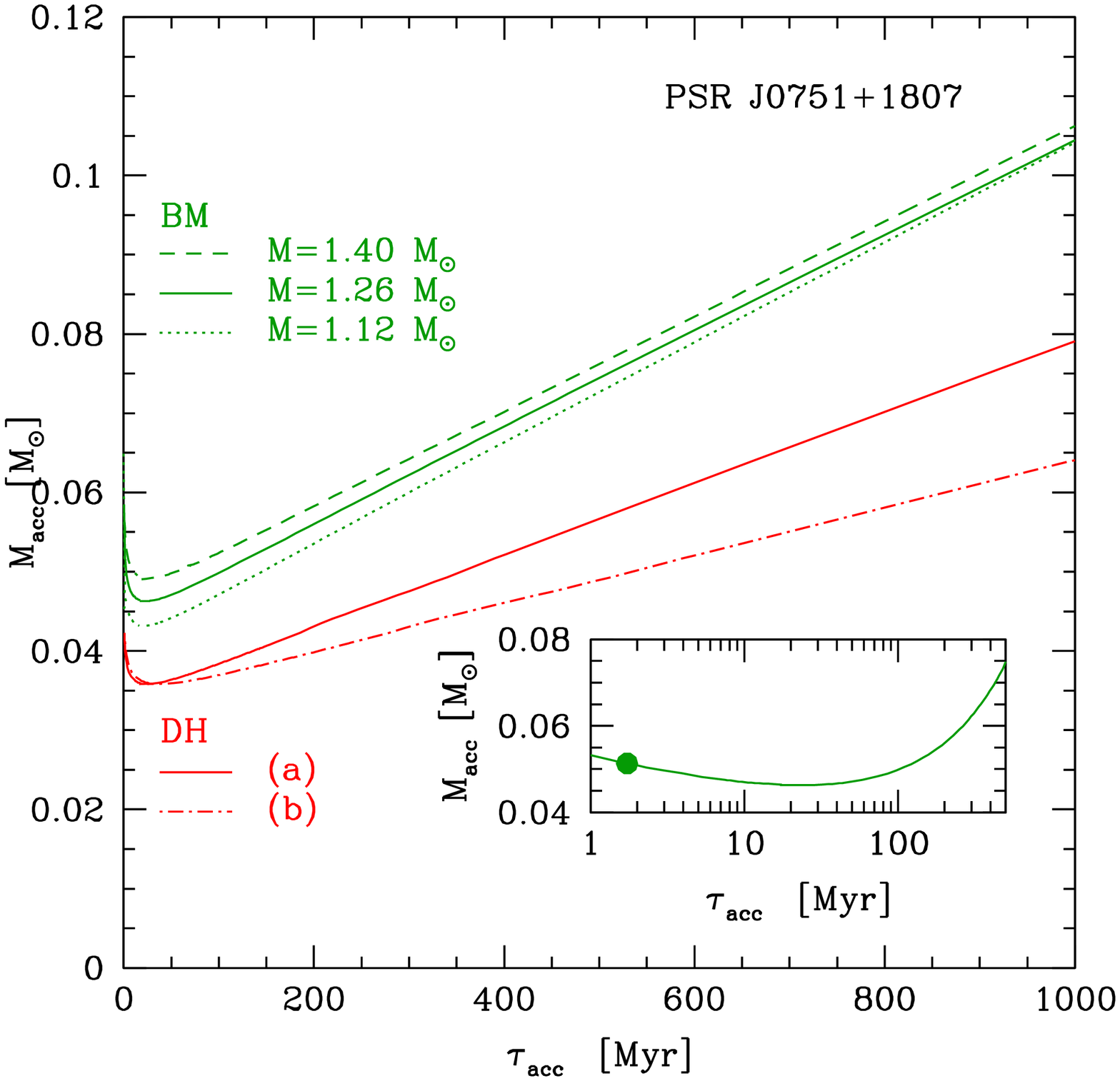}}
            \caption{(Colour online) Similar to Fig.~\ref{fig:tmacchigh}, but for the pulsar J0751+1807.}
            \label{fig:tmacclow}
\end{figure}

\begin{figure}
            \resizebox{\hsize}{!}{\includegraphics[scale=1]{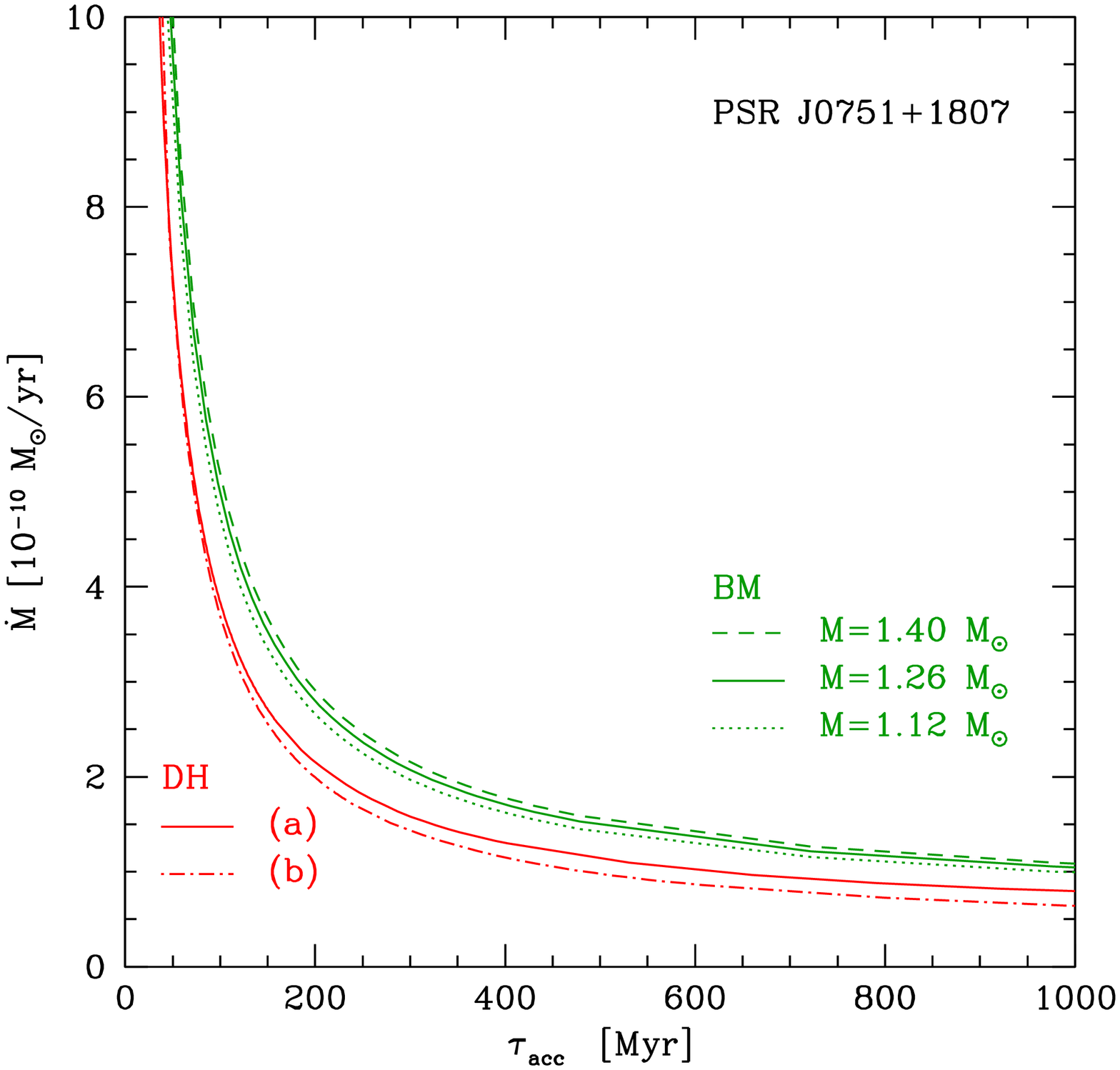}}
            \caption{(Colour online) Similar to Fig.~\ref{fig:mdotthigh}, but for the pulsar J0751+1807.}
            \label{fig:mdottlow}
\end{figure}

\begin{figure}
            \resizebox{\hsize}{!}{\includegraphics[]{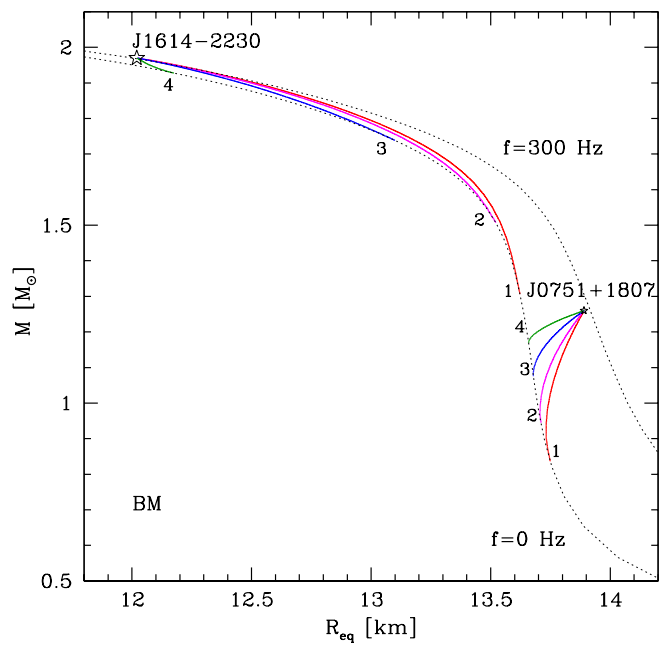}}
            \caption{(Colour online) Mass $M$ vs equatorial radius $R_{\rm eq}$ diagram and spin-up tracks of the two pulsars.
            The colour code and the numbers labelling the tracks correspond to the ones used in Figs.~\ref{fig:fmhigh}
            and \ref{fig:fmlow}. For comparison relations between the mass and the equatorial radius of
            equilibrium configurations are shown for $f=0$ and 300 Hz. The large open-star symbol corresponds to the PSR J1614$-$2230
             configuration and the small one to PSR J0751+1807.}
                \label{fig:mr}
\end{figure}

A similar approach was used for PSR J0751+1807. Properties of the five models are presented in Table \ref{tab:bmin}, and results are shown in Figs.~\ref{fig:fmlow}, \ref{fig:tmacclow}, and \ref{fig:mdottlow}. As before, spin-up tracks 1, 2, and 3 in Fig.~\ref{fig:fmlow} are plotted for illustrative purposes, and tracks 4 and $4^\prime$ are obtained for $m_{\rm B}=10^{-5}$, $10^{-4}~$M$_\odot$ respectively. The figures are remarkably similar to the ones obtained for PSR J1614$-$2230. As for PSR J1614$-$2230, the only difference between results for $m_{\rm B}=10^{-4}$ and $10^{-5}~$M$_\odot$ lies in the value of the initial magnetic field.

For PSR J1614$-$2230, however, the amount of accreted matter needed to spin
up the pulsar and the mean accretion rate decreases with increasing final mass of
the pulsar. This is visualized by the order of the curves corresponding to masses
$1.93,\, 1.97$,\, and $2.01~\msun$ in Figs.~\ref{fig:tmacchigh} and \ref{fig:mdotthigh}.
This behaviour is opposite to what is found for PSR J0751+1807
(Figs.~\ref{fig:tmacclow} and \ref{fig:mdottlow}).
The reason is the non-monotonic dependence of the moment of inertia $I$ (and total angular momentum of a star rotating at a fixed frequency)
on the stellar mass $M$ \citep{Bejger2013}. For the BM EOS, for example, for masses lower than $1.8\ \msun$, $I$ increases with $M$ and a higher mass corresponds
to a larger total angular momentum. As a consequence, for an equal transfer of
angular momentum by accretion, a longer time is needed to spin up the star to a given frequency. The situation is opposite close to the
maximum mass (for $M > 1.85~\msun$ for BM model) where $I$ decreases with $M$
(see Tables \ref{tab:bmax} and \ref{tab:bmin}).

As Fig.~\ref{fig:tmacclow} indicates, assuming that the accretion proceeds at a rate lower than the Eddington rate for $1~$Gyr (see Sect.~\ref{tab:197.126.numbers}), the progenitor NS of PSR J0751+1807 acquired $\sim 0.06-0.10\ \msun$, which implies that it was born with a very low mass, $\sim1.05-1.30\ \msun$.

 The wide range of possible birth masses for PSR J1614$-$2230 and PSR J0751+1807 puts stringent constraints on the modelling of supernova explosion and their outcome and on the EOS of hot dense matter, which is relevant for newly born NSs.

Figure~\ref{fig:mr} shows the spin-up tracks in the mass $M$ vs. equatorial
radius $R_{\rm eq}$ diagram for the two pulsars. During the recycling phase, the two NSs undergo
a remarkably different evolution in the $M-R_{\rm eq}$ diagram: the equatorial
radius of PSR J0751+1807 increases, while the one of PSR J1614$-$2230 substantially
decreases. The role of rotation at frequency $\sim 300$~Hz is much more important
for the structure and oblateness
 of a star with the mass $1.3\;\msun$ and radius $14$~km than for
 a much more compact NS with $M$ close to the maximum mass.
\section{The age of the MSPs and their true initial spin period}
\label{sect:spin-down}
The present spin period of MSPs is expected to be longer
than the period at the end of the spin-up phase. The pulsars' spin-down age can be estimated with the formula $\tau_{_{\rm PSR}}=P/2{\dot P}$. For the values given in Table \ref{table:psr}, one obtains $\tau_{_{\rm PSR}}^{\rm H}=5.2$~Gyr and $\tau_{_{\rm PSR}}^{\rm L}=7.7$~Gyr. These values should be
treated as upper limits on $\tau_{_{\rm PSR}}$ since the measured $\dot P$ is larger than the true $\dot P$ because of the transverse motion
of the binary system \citep{Shklovskii70}. This effect turns out to be negligible for the low-mass pulsar (about $\sim 1\%$). It may be crucial for the high-mass pulsar since the true ${\dot P}$ could be more than one order of magnitude smaller than the observed one \citep{Bhalerao2011}. However this result is very sensitive to the uncertainties in the determination of the distance and proper motion of PSR J1614$-$2230. Moreover, the spin-down age has been shown not to provide a reliable age estimate, in particular for MSPs, see e.g. \cite{KT10}.

The loss of the tenuous WD envelope marks the end of the LMXB stage, which leads to the formation of a WD+MSP binary. In both cases, L-binary with PSR J0751+1807 and H-binary with PSR J1614$-$2230, 
the heating of the WD surface by the MSP irradiation can be neglected. This is quite 
 natural for the H binary with $P_{\rm orb}=8.7~$d. It is far from being obvious 
 for the L-binary with $P_{\rm orb}=6.3~$h, but \cite{Bassa2006} found ``a surprising lack of evidence for any heating''. Therefore, the age of the MSPs in both the L and H binaries can be obtained from modelling the cooling of their WD.

Using their own observations of WD(L) \cite{Bassa2006} constrained its effective temperature $T_{\rm eff}$, radius, and the composition of its atmosphere. They found $T_{\rm eff}~\approx~4000~$K for the most likely pure He (or strongly He-dominated) atmosphere. Such a composition of the atmosphere is 
a puzzle (see discussion in \citealt{Bassa2006}). We applied a cooling curve obtained for a $0.15\;\msun$ He-core WD with a He envelope calculated by \cite{Hansen1998a} and obtained $\tau_{_{\rm WD}}^{\rm L}\approx 2.4~$Gyr. We assume that this is the age of PSR J0751+1807.

The WD companion of PSR J1614$-$2230 was discovered in the optical observations by \cite{Bhalerao2011} and in the following, we summarize their results. The optical colours, combined with a mass of $0.5~\msun$, indicate that the WD(H) has a C/O core and an H atmosphere. Cooling sequences calculated by \cite{Chabrier2000} predict that such a WD reaches the inferred absolute magnitude in the R-band $M_{\rm R}\approx 13.7$ obtained in \cite{Bhalerao2011} after $\tau_{_{\rm WD}}^{\rm H}\approx 2.2$~Gyr. This is the age we assume for PSR J1614$-$2230.

Using the measured $\dot P$ and assuming a braking index $n=3$, which is consistent with the observed population of MSPs \citep{KT10}, the initial period of a pulsar at the beginning of the slowing down phase can be estimated: $P_{\rm init}=P_{\rm obs}\left[1-\left(n-1\right)\tau_{_{\rm WD}}\dot{P}/P\right]^{1/\left(n-1\right)}$, where $P_{\rm obs}$ is the currently observed period. Here, $P_{\rm init}$ is shorter than $P_{\rm obs}$ by about $20\%$ for both pulsars. For the light pulsar, the amount of accreted matter necessary to spin it up not to $P_{\rm obs}$ but to $P_{\rm init}$, which is $\sim 30\%$ larger. As a consequence, its progenitor NS could be born with a mass $\sim 1.0\;\msun$.
\section{Discussion and conclusions}
\label{sect:conclusions}

We have presented the modelling of the accretion-induced spin-up phase undergone by two millisecond pulsars: the massive PSR J1614$-$2230 and the light PSR J0751+1807.
In agreement
with our previous work \citep{Bejger2011}, we showed that including the effect of the
NS magnetic field is crucial for a correct understanding of the
formation of millisecond pulsars. The dependence of our results on the
prescriptions used for the magnetic field accretion-induced decay, on the estimate of the
magnetic field, and on the EOS is remarkably small.

Our results indicate that the mass of the progenitor NS, born in a SNIb/c explosion, of PSR J1614$-$2230 was not lower
than $1.9~\msun$ and therefore was very close to the currently
measured value. This value is $\sim 0.2~\msun$ higher than the NS birth mass obtained in \citet{TaurisLK2011},
\citet{Tauris2012}, and \citet{LinRappaport2011}, in which the accretion of $M_{\rm acc}=0.3~\msun$ is required,
while for our spin-up models, $\sim0.06~\msun$ is
sufficient. This large discrepancy partly comes from differences in the modelling of the spin-up phase but mostly from the fact that we do not model the whole evolution of the binary system.

Equation\;14 in \citet{Tauris2012} yields a minimum value of $M_{\rm acc}$. This value 
is remarkably similar to our value of $M_{\rm acc}$. 
Our calculations show that this estimate is independent of the EOS and of the assumed model for the magnetic field and
its accretion-induced decay.
A reasonable estimate of the accreted mass could be obtained on the basis of the current parameters of the pulsar
($P$, $\dot P$, $M$). In our approach the evolution depends on the mean accretion rate, which is not assumed to be
the equilibrium one. However, there is a minimum value of the accreted mass $M_{\rm acc}^{\rm min}$ needed to spin the star up to the observed rotational
period. In our model, $M_{\rm acc}^{\rm min}\simeq 0.05\ \msun$ (for details see Appendix) and is obtained for a given $\tau_{\rm acc}^{\rm min}$ (see e.g. Figs.~\ref{fig:tmacchigh} and \ref{fig:tmacclow}). The mean accretion rate is then given by $\dot{M}^{\rm min}=M_{\rm acc}^{\rm min}/\tau_{\rm acc}^{\rm min}$.
All these values for the amount of accreted matter
are actually lower limits since processes in the NS magnetosphere and in the accretion disk at the origin
of the ejection are not included.

In our approach \citep{Bejger2011}, relativistic effects are taken into account by introducing dimensionless function $f_{ms}$ in Eq.\;(\ref{bc}).
The role of GR for the radius $r_0$ is negligible, of the order of 1\% (for details see Appendix in
\citealt{Bejger2011}). However, the specific angular momentum of a particle $l(r_0)$ calculated in the framework of GR is larger
than obtained in Newtonian theory. The increase in $l$ due to GR is maximal
at a marginally stable orbit (by a factor $\sqrt{2}$, see \citealt{kluwag}), and in cases considered in this paper ($r_0>\rms$)
is $\sim 10\%$. As a consequence, disregarding GR effects results in less effective spin-up;
the mass needed to reach given frequency is therefore higher by $\sim 10\%$.

The progenitor of PSR J0751+1807 was itself born with a very
low mass, which as we estimate, could be as low as $1.05\;\msun$. 
Considering that the pulsar spun down after the recycling, the mass of the progenitor NS is lowered to $1.0\;\msun$.

The Roche
lobe decoupling phase (RLDP), suggested recently by
\citet{Tauris2012Science} and \citet{Tauris2012} is related to some additional quasi-spherical
accretion. Although we did not model the RLDP, we estimated the
additional $M_{\rm acc}({\rm RLDP})$ as follows: When assuming a slow RLDP with a
timescale of $\simeq 50$ Myr and a pre-RLDP accretion rate
$\dot{M}\sim 10^{-10}\ \msun/{\rm yr}$ in the case of PSR
J0751+1807, the upper limit on $M_{\rm acc}({\rm RLDP})$ is 0.005 $\msun$.
Because of the rapidly decreasing
$\dot{M}$ and because some matter
is ejected during the propeller phase, $M_{\rm acc}({\rm RLDP})$ is likely to be smaller. Considering
that before the RLDP phase, a pulsar should have been spun up to a
higher frequency than the one observed now, by accreting more mass
before this phase, we can therefore conclude that the amount of matter accreted during the RLDP 
is negligible compared to the total mass accreted by the NS during the recycling process.

The wide range of NS birth masses, $1.0~\msun - 1.9~\msun$ derived from our simulations, agrees with recent modellings by \citet{UJ12} and \citet{PT14} of supernova explosions of a large set of massive stars
progenitors and metallicities in spherical symmetry.

\begin{acknowledgements}
{We are grateful to Antonios Manousakis, Janusz Zi{\'o}{\l}kowski,
Hans-Thomas Janka, and an anonymous referee for reading the
manuscript and for helpful remarks and suggestions. We also
acknowledge helpful remarks of participants of the CompStar 2011
Workshop (Catania, Italy, 9-12 May, 2011). This work was
partially supported by the Polish NCN research grant no.
2013/11/B/ST9/04528 and by the COST Action MP1304 ''NewCompStar''.}
\end{acknowledgements}


\section*{Appendix: Approximate solution}
\label{sect:appendix}

The properties of the solutions of Eqs.\;(\ref{eq:ltot}) and (\ref{bc})
allow us to determine an approximate solution for our model. It contains only
the parameters of the final configuration: $B$, $f$, $M$ and a given mean
accretion rate $\dot M$ and enables the properties of the pre-accretion NS to be estimated. 
The spin-up equation (\ref{eq:ltot}) can be factorized as
\begin{equation}
 \frac{{\rm d} J}{{\rm d} M_b}=\sqrt{GMr_c}\cdot \lambda(f)
,\end{equation}
where $\lambda\equiv {\ltot}/{\sqrt{GMr_c}}$ is the ratio of total
specific angular momentum of an accreted particle to its value at
the corotation radius. The main dependence on rotational frequency is
included in $\sqrt{r_c}\sim f^{-1/3}$.

The ratios of the characteristic lengths describing the system ($r_m$, $r_c$, $r_0$)
do not change significantly along a spin-up
track, except for a small region of very
 slow rotation. Therefore, one can assume that $\lambda (f)$ is constant along
 spin-up track as the Eq.\;(\ref{bc}) depends on $r_m/r_0$ and $r_c/r_0$.
 This assumption holds with an accuracy of about 20\%.
 However, $\lambda$ is a rather sensitive function of the mean accretion rate.
 For example, for the set of evolutionary tracks with different accretion rates
 presented in Figs.~\ref{fig:evol197} and \ref{fig:evol126}, the
 value of $\lambda$ changes by more than one order of magnitude.

 Neglecting the change in the NS moment of inertia $I$ (which is a good assumption
 for configurations close to the maximum mass and/or for
relatively small amount of accreted mass), we obtain the formula
\begin{equation}
 \Delta M \simeq \frac{3}{4}\frac{J_{f}}{l_{f}},\quad{\mathrm{with}\quad}l_f=l(r_{0f})-\lmag,
\end{equation}
where $J_f$ is the angular momentum, $l(r_{0f})$ is specific
angular momentum of an accreted particle, and $\lmag$ is magnetic
torque divided by $\dot M$, all three quantities corresponding to the final (presently
observed) state of the pulsar.

\begin{figure}
            \resizebox{\hsize}{!}{\includegraphics[]{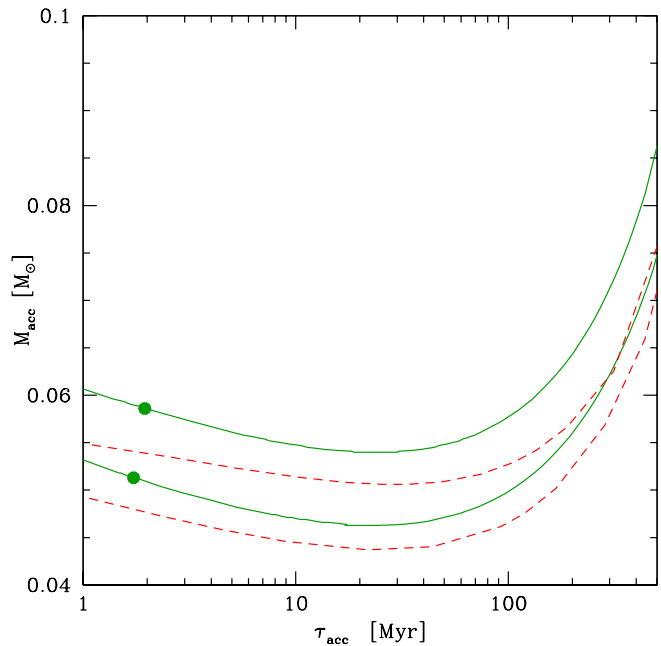}}
            \caption{(Colour online) Mass of accreted matter as a function of the time needed to spin-up
            the progenitor NS to the observed state (i.e. $M$, $B,$ and $f$) for PSR J1614$-$2230 (upper curves)
            and PSR J0751+1807. Exact results - solid (green) curves; approximation - dashed (red) lines.}
             \label{fig:approx}
\end{figure}

Since the magnetic torque is proportional to $1/{\dot M}=\tau/{\Delta
M}$ in the limit of a low accretion rate (large $\tau$), the angular momentum transferred
to the star is $l_{f}=l(r_0)\cdot~(1~-~A/{\dot M})=l(r_0)\cdot (1 - A\tau/{\Delta M})$, the
dependence on $\tau$ is almost linear, as shown in Figs.~\ref{fig:tmacchigh} and \ref{fig:tmacclow}:
\begin{equation}
 \Delta M \simeq \frac{3}{4}\frac{J_{f}}{l_0}+A\cdot \tau,\quad{\mathrm{with}\quad 
 A=\frac{\mu^2}{9r_0^3 l(r_0)}
 \left(3-2\sqrt{\frac{r_c^3}{r_0^3}}\right)}.
\label{eqn:approx}
\end{equation}
A simple estimate of the amount of accreted matter needed to spin up a NS to a given configuration can be obtained from Eq.\;(\ref{eqn:approx}). The radius $r_{0f}$
at the inner boundary of the disk can be calculated by solving the simple algebraic equation Eq.\;(\ref{bc})
for values of $f$, $B$, and $M$ corresponding the ones of the final configuration. Then $l(r_{0f})$ can be
determined by the analytic formula given in \cite{Bejger2010}.
Examplary results based only on the present-day parameters are shown in Fig.~\ref{fig:approx}.
For a given duration of the accretion phase, one can then derive the amount of matter accreted by the NS to reach its current configuration. Then the birth mass can be simply derived and the pre-accretion magnetic field is given by Eq.\;(\ref{eq:B.DeltaM}).

For high $\dot M$, the value of $A$ depends sensitively on the
solution of Eq.\;(\ref{bc}), and the function $A(\dot M)$ changes
its sign at the point $\dot{M}^{\rm min}$ corresponding to
\begin{equation}
\left(\frac{r_0}{r_c}\right)^{3/2}=\frac{2}{3},\quad{\rm
i.e.,}\quad r_c=1.31\, r_0.
\end{equation}
For $\md>\dot{M}^{\rm min}$, the accreted mass $\Delta M$ is a decreasing function
of $\tau_{\rm acc}$ (see inserts in Figs.~\ref{fig:tmacchigh} and \ref{fig:tmacclow}). 
In our case, $\dot{M}^{\rm min}\simeq 2\times 10^{-9}\ \msolyr$ and thus is more than one
order of magnitude lower than the Eddington limit $\simeq 3\times
10^{-8}\,\msolyr$. The minimum mass needed to spin a 
pulsar up to its observed frequency depends on the value of the specific angular momentum
of a particle at the corotation radius and is given by
$\Delta M^{\rm min} = 0.75 J_{f}/l(0.763\, r_c)$. This quantity does not depend on the accretion rate.

\end{document}